\documentclass[12pt]{article}

\usepackage{amsmath}
\usepackage{amssymb}

\newcommand{\eps}{\varepsilon}
\newcommand{\ga}{\gamma}
\newcommand{\om}{\omega}

\begin{document}

\title{Theory of the radiative width of a highly excited nucleus
\footnote{Zh. Eksp. Teor. Fiz. {\bf 67}, 441--452 (1974) [Sov. Phys. JETP
{\bf 40,} No.~2, 219-224 (1975)]}}

\author{V.G. Nosov$^{\dagger}$ and A.M. Kamchatnov$^{\ddagger}$\\
$^{\dagger}${\small\it Russian Research Center Kurchatov Institute, pl. Kurchatova 1,
Moscow, 123182 Russia}\\
$^{\ddagger}${\small\it Institute of Spectroscopy, Russian Academy of Sciences,
Troitsk, Moscow Region, 142190 Russia}
}

\maketitle

\begin{abstract}
A microscopic theory of the electromagnetic radiation emitted by a highly excited
nucleus is developed on the basis of the Landau theory of a Fermi liquid.
Closed formulae are obtained for the mean radiative width and its mean square
fluctuation from level to level. The temperatures of many nuclei are found from
the observed widths. The relaxation
time is estimated from the experimental data on the radiative-width fluctuations.
The regions of applicability of the various types of relations between the
relaxation time and the lifetime of the compound nucleus, as well as the relevant
physical consequences, are discussed.
\end{abstract}

\section{Introduction}

That there is a gradual unification of the mechanisms underlying the emission
of electromagnetic radiation by a not too light nucleus as the excitation energy of
the nucleus increases is practically beyond question. Roughly speaking, if an
infinite number of the levels of the nucleus as a whole lie below the initial
excitation energy, then the system will itself find and prefer the easiest effective
way of emitting $\ga$ quanta. In particular, there is below the initial excited
level an abundance of levels $\ga$ transitions to which satisfy the most favorable
``selection rules", so that from this point of view the process is, in the limit
under consideration, practically one with an infinite number of channels. Analysis
shows (see [1]) that the electric-dipole radiation due to collisions between
the (proton and neutron) quasiparticles and the ``wall" of the nucleus predominates.
The ideas of the Landau theory of the Fermi fluid [2,3] allow us to compute in closed
form the radiative width $\overline{\Gamma}_\gamma$ and its fluctuations from level to level.

In a sense, spherical nuclei are rather exotic objects:
the application to them of the Fermi-fluid concepts requires in each specific case
certain precautions. The fact that the spherical configuration is stable is in itself
an indication of the essential role played by the ``residual interaction" between the
quasiparticles, an interaction which blurs the Fermi level: it can be shown that in
the scheme without interaction the sphere is absolutely unstable (see [4]). Furthermore,
analysis of the data on the shell and magic oscillations in the masses of spherical
nuclei allows the establishment of the macroscopically ordered structure that this
residual interaction possesses in the space of the values of the orbital momentum $l$
of the individual quasiparticles [5]. (We shall again touch upon spherical
nuclei when we compare below the theoretical results with the experimental data.)
Nonspherical nuclei, on the other hand, are easy to investigate, it being apparently
necessary to regard their shape as a perfectly natural consequence of the properties
of the ``normal", disordered nuclear phase, in which the quasiparticles situated near
the limit of the Fermi distribution move, in the main, independently of each other.

However, for the theory of radiative widths expounded below the ``shape effects", as such,
are of no independent importance, and a special allowance for them is not necessary.
Indeed, the equilibrium deformation $\alpha$ of a nonspherical nucleus is equal in order of
magnitude to $\rho_f^{-1}$ (i.e., $\alpha\sim\rho_f^{-1}$), where
\begin{equation}\label{1}
    \rho_f=k_fR\gg 1
\end{equation}
($k_f$ is the limiting momentum of the quasiparticle distribution and $R$ is the radius
of the nucleus) is an important dimensionless parameter that arises in the most diverse
investigations that have as their aim the treatment of the nucleus as a macroscopic body.
In view of the scalar nature of the quantity $\Gamma_\gamma$ to be computed, only the
squared deformation can enter, so that the relative magnitude $\alpha^2\sim\rho_f^2\ll 1$
of the corresponding corrections is negligibly small and falls outside the limits of
accuracy of the theory. In other words, only the possibility of considering the
quasiparticles individually is important for what follows, it still being possible in
actually occurring deformations to treat the geometry of the motion of each of the
quasiparticles as spherically symmetric.

\section{The radiative width of a highly excited nuclei}

We shall derive the expressions, referred to one quasiparticle, on the basis of the
correspondence principle. As applied to radiation processes, this principle asserts
total analogy between the formulas of the classical and quantum theories (see, for
example [6]). The classical intensity $I$ (i.e., the energy emitted per unit time)
needs only to be divided by $\eps=\hbar\om$ to be converted into the quantity of real
interest---the probability of emission of individual $\ga$ quantum. The only remaining
difference consists in the following: the spectral component of the multipole moment,
which varies according to a classical law, should, generally speaking, be replaced by
the corresponding matrix element of its operator. They, however, coincide in the
quasiclassical limit (see (1)) [7]. Consequently, we can speak of a quasiparticle
trajectory: it is in this case a straight line joining two points on the
surface of the nucleus.

The basic formula of the classical theory of the electric dipole radiation has the form
\begin{equation}\label{2}
    I=\frac{2e_d^2}{3c^3}\,\ddot{\mathbf{r}}^2,
\end{equation}
where $e_d$ is the charge of the radiating particle (the radiating quasiparticle,
in the general case; see below) and $\ddot{\mathbf{r}}$ is its acceleration vector.
Before proceeding to the spectral decomposition of $I$, let us note that in the
thermal-equilibrium state the quasiparticle motion inside the nucleus does not change
its qualitative character in time. Therefore, let us formally carry out the Fourier
expansion over an arbitrary, but sufficiently long interval of time $t$ and then
take the limit as $t\to\infty$ (see [8]):
\begin{equation}\label{3}
    \overline{\ddot{\mathbf{r}}^2}=\frac1{\pi t}\int_0^\infty\left|\int_0^t
    \mathbf{r}(t)e^{i\om t}dt\right|^2\om^4d\om.
\end{equation}
In view of the independence of the different chords traced by the quasiparticles in
their wall-to-wall motion, we have
\begin{equation}\label{4}
    \left|\int_0^t\mathbf{r}e^{i\om t}dt\right|^2=\overline{\left|\int_0^{t_1}
    \mathbf{r}e^{i\om t}dt\right|^2}\,n \cong \overline{\left|\int_0^{t_1}
    \mathbf{r}dt\right|^2}\,n,
\end{equation}
where $t_1 = l/v_f$ is the time it takes to travel from one end of a chord to the
other, $l = 2\sqrt{R^2 -\rho^2}$ is the length of the chord, and $n$ is the number
of chords. Here we have taken into account the fact that in the region of the
radiation energy spectrum of interest to us
\begin{equation}\label{5}
    \om t_1\ll 1
\end{equation}
(we shall return to the criterion (5) later). Furthermore, here and below the
quasiparticle velocity $v$ is replaced everywhere by its limiting value $v_f$.
The point is that because of the Pauli principle only those ``elementary emitters"
(i.e., quasiparticles) that are situated in the immediate neighborhood of the Fermi
level play a role (see below).

The distribution of the chords over the impact parameters $\rho$ is easily found
from considerations of isotropy and homogeneity of nuclear matter:
\begin{equation}\label{6}
    w(\rho)d\rho=\frac3{R^3}\sqrt{R^2-\rho^2} \rho d\rho;\quad
    \int_0^Rw(\rho)d\rho=1.
\end{equation}
Averaging, in accordance with (4) and (6), the square of the integral over the
radius vector, we also express the number $n$ of quasiparticle---nuclear wall
collision events in terms of the physical time $t$:
\begin{equation}\label{7}
    \overline{\left|\int_0^{t_1}\mathbf{r}dt\right|^2}=\frac4{v_f^2}
    \overline{\rho^2(R^2-\rho^2)}=\frac{24}{35}\frac{R^4}{v_f^2},\quad
    n=t\frac{v_f}{\bar{l}}=\frac23\frac{v_f}{R}t.
\end{equation}
Taking into consideration the relations (2)-(7) and the relevant considerations,
we obtain
\begin{equation}\label{8}
    f(\eps)d\eps=\frac{32}{105\pi}\frac{e_d^2}{\hbar^5c^3}\frac{R^3}{v_f}\eps^3d\eps.
\end{equation}
This expression gives the probability per unit time of emission by one quasiparticle
of a $\ga$ quantum in the interval $d\eps$ of its energy values.

According to the theory of the Fermi fluid [2-3], the mean occupation numbers of the
individual quantum states are given by the standard Fermi distribution
\begin{equation}\label{9}
    \bar{n}(\eps')=\frac1{e^{\eps'/T}+1},
\end{equation}
where $\eps'$ is the quasiparticle energy measured relative to the chemical potential
and $T$ is the temperature. On the other hand, the number of actual single-quasiparticle
states in the volume $V = (4/3)\pi R^3$ is equal to
\begin{equation}\label{10}
    d\widetilde{N}=\frac{V}{\pi^2\hbar^3}\frac{p^2}{d\eps'/dp}d\eps'\cong
    \frac{4R^3}{3\pi\hbar^3}\frac{p_f^2}{v_f}d\eps'
\end{equation}
($p_f =\hbar k_f$ is the limiting momentum in standard units), where allowance has
been made for the additional spin doubling. In fact, even in the quasiclassical limiting
case, (1) remains an important quantum effect due to the identity, the
indistinguishability of identical fermions [7]: the above-described classical picture
of the process is actually realizable only in the case of radiative transitions that
are compatible with the Pauli principle. Therefore, the product of the expressions
(8), (9), and (10) should be supplemented by the factor
$$
1-\bar{n}(\eps'-\eps),
$$
which determines the fraction of the transitions admissible by this principle.
Then integration over the energies $\eps'$ of the radiating quasiparticles reduces to
\begin{equation}\label{11}
    \int_{-\infty}^\infty\bar{n}(\eps')[1-\bar{n}(\eps'-\eps)]d\eps'=
    \frac\eps{e^{\eps/T}-1}.
\end{equation}
Integration over the boson energies yields
\begin{equation}\label{12}
    \int_0^\infty\frac{\eps^4d\eps}{e^{\eps/T}-1}=24\zeta(5)T^5,\quad
    \zeta(z)=\sum_{n=1}^\infty\frac1{n^z},
\end{equation}
where $\zeta$ is the Riemann zeta function. Finally, the $\ga$-quantum emission
probability per unit time, appropriately summed over the entire set of quasiparticles
of the same sort will be given by
\begin{equation}\label{13}
    W=\frac{1024}{105\pi^2}\zeta(5)\frac{e_d^2m^{*2}}{\hbar^8c^3}R^6T^5,
\end{equation}
where $m^* = p_f/v_f$ is the effective mass of the quasiparticle.

Above, as the coordinate origin convenient for the calculations, we used the
geometrical center of the nucleus. However, the role of the total charge $Ze$ of the
whole system in processes induced by the oscillations of the radius vector of the
individual nucleons (quasiparticles) is well known. Because of recoil, even the
electrically neutral quasiparticles (i.e., the neutron quasiparticles) will appear
to emit radiation during their motion relative to the center of the nucleus. The
corresponding, well-known, ``charge-renormalization" formulae have the form
\begin{equation}\label{14}
    e_d^Z=\left(1-\frac{Z}A\right)e,\quad e_d^N=-\frac{Z}A e
\end{equation}
(the ``effective charges," (14), of the two components are correct only for processes
induced by the oscillations of the electric dipole moment of the nuclear system
(see, for example, [7])). Summing, with allowance for (14), the expressions (13)
or the proton and neutron components of the nuclear matter, and multiplying them by
$\hbar$ in order to convert them into the energy widths of interest, we finally obtain
\begin{equation}\label{15}
    \overline{\Gamma}_\gamma=\frac{1024}{105\pi^2}\zeta(5)\frac{e^2m^{*2}}{\hbar^7c^3}
\left[1-2\frac{Z}A\left(1-\frac{Z}A\right)\right]R^6T^5
\end{equation}
(notice that the numerical factor $(1024/105\pi^2)\zeta(5)\cong 1$ is very close to unity).
The law $\overline{\Gamma}_\gamma\propto T^5$ was given in the preceding paper (see [1],
formula (4)), where it was motivated by semi-phenomenological considerations.

To what extent can the result (15) be identified with the radiative widths of the
individual resonance levels of a compound nucleus that is excited, say, in a reaction
involving slow-neutron capture? It follows from its derivation that the formula (15)
 corresponds to a state in which at the temperature $T$ the quasiparticles of the
 nuclear Fermi liquid are in thermal equilibrium with each other. On the other hand,
 the width $\Gamma_\gamma$  of a specific level can, reasoning abstractly, be
 conceived to have been computed from some very complicated, unknown (to us) wave
 function of the corresponding state of the nucleus as a whole. According to the
 fundamental principles of statistics, the two approaches lead to results that
 coincide to within the values of the fluctuations (see, for example, [3]).

Let us now rewrite the condition (5) of applicability of the theory in a more
concrete form. Owing to the thermal nature of the radiation, the inequality (5) is
equivalent to the following inequality:
\begin{equation}\label{16}
    T\ll\eps_0,
\end{equation}
where
\begin{equation}\label{17}
    \eps_0=\hbar v_f/2R\sim 5 \mathrm{MeV}
\end{equation}
is the characteristic energy corresponding to the reciprocal of the time it takes
a quasiparticle to cross the nucleus along a diameter.

It is worth noting that in the opposite limiting case
\begin{equation}\label{18}
    T\gg \eps_0,
\end{equation}
because of the oscillations of the exponent $e^{i\om t}$ along the chord traced by
the quasiparticle (see (3) and (4)), the energy distribution of the $\gamma$-quantum
emission probability acquires the form of the well-known Planck black-body radiation
spectrum [3]. The radiation width would, accordingly, become proportional to the cube
of the temperature in the case of a sufficiently strict fulfilment of the condition
(18). However, this ``black-body radiation limit" defined by (18) is, in practice,
hardly attainable in nuclear physics. At least the temperature of the compound nucleus
should not exceed the nucleon binding energy, which is  $\sim 8$ MeV---otherwise the
neutrons would fly out of the nucleus ``instantly," escaping the thermal-equilibrium
establishment phase \footnote{Notice that the radiation due to collisions between the
quasiparticles would then become dominant only at $T\gg\sqrt{\rho_f\eps_0}$.}.

\section{The mean-square fluctuation in the radiation width. The role of the
relaxation time}

The direct, quantum-mechanical computation of the characteristics of the individual
states of the nucleus is inexpedient and practically impossible. Furthermore, as
applied to macroscopic bodies (see the criterion (1)), this, as a rule, borders on
the theoretical impossibility [3]. Therefore, we are obliged here to treat the state
of the occupation of the individual quantum states of the quasiparticles of the Fermi
liquid as a randomly varying function of the time. We shall calculate the instantaneous
``emissive power" $\widetilde{\Gamma}_\gamma$ of the nucleus in a manner completely
similar to the computations of the preceding section. The ``one-component" variant of
the corresponding formula can be represented in the form
\begin{equation}\label{19}
    \widetilde{\Gamma}_\gamma=\frac{32}{105\pi}\frac{e_d^2}{\hbar^4c^3}\frac{R^3}{v_f}
    \int_0^\infty d\eps_1\cdot\eps_1^3\sum_{\eps'}n(\eps')[1-n(\eps'-\eps_1)].
\end{equation}
Here we have, for simplicity and convenience , written the discrete sum $\sum_{\eps'}$
over the fermion states. In case of need the transition to integration can easily be
accomplished with the aid of (10).

The ``instantaneous", physically realizable values
\begin{equation}\label{20}
    n_{\eps'}=0,\,1
\end{equation}
of the fermion occupation numbers differ from the mean occupation numbers (9). This
circumstance is the obvious cause of fluctuations in Fermi systems. It is convenient
to consider them with the aid of the simple relation (see [3])
\begin{equation}\label{21}
    \overline{\Delta n'\Delta n''}= \bar{n}'(1-\bar{n}')\delta_{\eps',\eps''}.
\end{equation}
Let us find the mean-square fluctuation of the expression (19)---the number of
summations and integrations doubles upon squaring. One summation over the
quasiparticle states is trivial owing to the presence of the $\delta$ symbol on the
right-hand side of (21); the subsequent integration is elementary, although
somewhat tedious. Adding, in accordance with (14), the squares of the fluctuations
in the proton and neutron components and introducing the dimensionless variables
$x_{1,2}=\eps_{1,2}/T$ in place of the $\ga$-quantum energies, we obtain
\begin{equation}\label{22}
    \begin{split}
    \overline{(\Delta\widetilde{\Gamma}_\ga)^2}=\frac{8192 J}{33075\pi^3}
    \frac{e^4m^{*3}}{\hbar^{12}c^6}\left[\frac{(1-Z/A)^4}{\rho_f^Z}+
    \frac{(Z/A)^4}{\rho_f^N}\right]R^{10}T^9,\\
    J=\iint_0^\infty\left[\frac{x_1e^{x_1}\coth(x_1/2)}{(e^{x_1}-e^{x_2})(e^{x_1}-e^{-x_2})}
    +\frac{x_2e^{x_2}\coth(x_2/2)}{(e^{x_2}-e^{x_1})(e^{x_2}-e^{-x_1})}\right]
    x_1^3x_2^3dx_1dx_2.
    \end{split}
\end{equation}
The details of the integration over the boson energies are given in the Appendix. The
final result has the form
\begin{equation}\label{23}
    J=\frac{848}{1575}\pi^8+576\sum_{n=1}^\infty\frac1{n^4}\sum_{k=n+1}^\infty
    \frac{k-n}{k^5}.
\end{equation}
Notice that the term with the double sum is about half percent of the value of the integral,
so that in practice we can restrict ourselves to the consideration of only the first term
on the right-hand side of (23).

The problems pertaining to the fluctuations are relatively subtle and require a more
careful physical treatment. In particular, there is no reason to equate
$\overline{(\Delta\widetilde{\Gamma}_\ga)^2}$ to the mean square
$\overline{(\Delta\Gamma_\ga)^2}$ of the actually observable, physical fluctuation in
the radiative widths of many close resonance levels. This becomes especially apparent
when we consider the most important and interesting case in which thermal equilibrium
in the nucleus is established long before the ``decay" of the nucleus:
\begin{equation}\label{24}
    \Gamma\tau/\hbar\ll 1.
\end{equation}
Here $\tau$ is the relaxation time (see below) and $\Gamma$ is the total width of the
initial state of the nucleus. Taking into account the fact that this quasistationary
state decays according to the law $\exp(-\Gamma t/\hbar)$, we express the number
$\nu$ of emitted quanta and its fluctuation in terms of the instantaneous emissive power
$\widetilde{\Gamma}_\ga$:
\begin{equation}\label{25}
    \begin{split}
    \nu& =\frac1\hbar\int_0^\infty\widetilde{\Gamma}_\ga(t)e^{-\Gamma t/\hbar}dt,\\
    (\Delta\nu)^2& =\frac1{\hbar^2}\iint_0^\infty\Delta\widetilde{\Gamma}_\ga(t)
    \Delta\widetilde{\Gamma}_\ga(t')\exp\left[-\Gamma(t+t')/\hbar\right]dtdt',\\
    \Delta\widetilde{\Gamma}_\ga(t)&=\widetilde{\Gamma}_\ga(t)-\overline{\Gamma}_\ga.
    \end{split}
\end{equation}
Further, it is convenient to introduce the notation $t' = t + \tau$. According to the
thermodynamic theory of non-equilibrium processes (and of the corresponding fluctuations
in the thermal-equilibrium state; see, for example, [3]), the mean value of the time
correlation of the fluctuations is given by the relation
\begin{equation}\label{26}
    \overline{\Delta\widetilde{\Gamma}_\ga(t)\Delta\widetilde{\Gamma}_\ga(t+\tau)}=
    \overline{(\Delta\widetilde{\Gamma}_\ga)^2} \,\exp(-|\tau|/\bar{\tau}),
\end{equation}
where $\bar{\tau}$ is the relaxation time. With allowance for (24), the
substitution of (26) into (25) yields
\begin{equation}\label{27}
    \overline{(\Delta\nu)^2}\cong\frac{\overline{(\Delta\widetilde{\Gamma}_\ga)^2}}{\hbar^2}
    \int_0^\infty dt\exp\left(-\frac{2\Gamma}\hbar t\right)\int_{-\infty}^\infty
    \exp\left(-\frac{|\tau|}{\bar{\tau}}\right)d\tau=
    \frac{\overline{(\Delta\widetilde{\Gamma}_\ga)^2}}{\hbar\Gamma}\bar{\tau}.
\end{equation}

Let us now consider the ensemble of the large number of close levels of a compound
nucleus with radiative width $\Gamma$: owing to the fact that the levels decay according
to the single law $\exp(-\Gamma t/\hbar)$, the equilibrium in the ensemble (the
equipopulation of the levels) is not destroyed in time. The number of $\ga$ quanta
$$
\nu=\frac1\hbar\int_0^\infty\Gamma_\ga e^{-\Gamma t/\hbar}dt=\frac{\Gamma_\ga}{\Gamma}
$$
has been pre-averaged over a group consisting of many levels with practically the same
$\Gamma_\ga$. In the final averaging of the square of the fluctuations $(\Delta\nu)^2$
over the entire ensemble of the groups differing in their radiative widths $\Gamma_\ga$,
each group is taken into account with a weight proportional to the number of levels in it:
\begin{equation}\label{28}
    \overline{(\Delta\nu)^2}=\overline{(\Delta\Gamma_\ga)^2}/\Gamma^2.
\end{equation}
Equating the right-hand sides of the formulas (27) and (28), we finally obtain
\begin{equation}\label{29}
    \overline{(\Delta\Gamma_\ga)^2}=(\Gamma\tau/\hbar)\overline{(\Delta\Gamma_\ga)^2}.
\end{equation}
(we shall no longer write the averaging sign over the relaxation time $\tau$). A striking
feature of the relation (29) consists in the following: It turns out that the fluctuations
in the probability of decay of the compound nucleus via the radiative channel depend on
the total decay probability $\Gamma$, including all the generally possible decay channels.
The physical meaning of the formula (29) is simple: In the time picture the deviation of
$\Delta\overline{\Gamma}_\ga(t)$, the emissive power, from its mean value has time to
average out to some extent provided the decaying exponential function varies sufficiently
slowly (see the criterion (24)). The small factor $\Gamma\tau/\hbar$ on the right-hand
side of (29) is precisely the quantity that determines the fraction of the physical,
actually observable effect that remains after such a partial averaging.

\section{Comparison with experiment}

With the aid of the formula (15) we determined the temperatures of compound nuclei from
the observed radiative widths of their resonance levels [9,10]. The results of such an
analysis for two well-known regions of nonspherical nuclei are given in the tables. We
assumed in the computations that
\begin{equation}\label{30}
    R=1.2\cdot 10^{-13}A^{1/3}\, \mathrm{cm}
\end{equation}
and $m^* = m_n$, where $m_n$ is the mass of the free nucleon. It is noteworthy that
the temperature in the case of the actinide nuclei turns out consistently to be
$\sim 100$ keV lower than the characteristic temperature for the lanthanide region.
This may be due to both the decrease of the neutron attachment energy toward the end
of the Mendeleev periodic table and the difference in the atomic weight $A$. A similar
temperature decrease apparently occurs within the nonspherical-lanthanide region.

\begin{table}
\begin{tabular}{|c|c|c|c|}
\hline
Compound nucleus & $E_{max}$, MeV & $\bar{\Gamma}_\ga$, MeV &$T$, MeV \\
\hline
\multicolumn{4}{|c|}{Nonspherical lanthanides} \\
$_{62}$Sm$_{86}^{148}$ & 8.14 & 52 & 0.42  \\
$_{62}$Sm$_{88}^{150}$ & 7.98 & 64 & 0.44  \\
$_{62}$Sm$_{91}^{153}$ & 5.89 & 71 & 0.45  \\
$_{63}$Eu$_{89}^{152}$ & 6.29 & 89 & 0.47  \\
$_{63}$Eu$_{91}^{154}$ & 6.39 & 102 & 0.48  \\
$_{64}$Gd$_{92}^{156}$ & 8.53 & 110 & 0.48  \\
$_{64}$Gd$_{93}^{157}$ & 6.35 & 110 & 0.48  \\
$_{64}$Gd$_{94}^{158}$ & 7.93 & 89 & 0.46  \\
$_{64}$Gd$_{95}^{159}$ & 6.03 & 105 & 0.48  \\
$_{65}$Tb$_{95}^{160}$ & 6.40 & 90 & 0.47  \\
$_{66}$Dy$_{96}^{162}$ & 8.20 & 122 & 0.49 \\
$_{66}$Dy$_{97}^{163}$ & 6.25 & 175 & 0.52 \\
$_{66}$Dy$_{98}^{164}$ & 7.66 & 103 & 0.47 \\
$_{66}$Dy$_{99}^{162}$ & 5.64 & 166 & 0.51 \\
$_{66}$Dy$_{96}^{162}$ & 8.20 & 122 & 0.49 \\
$_{67}$Ho$_{99}^{166}$ & 6.33 & 91 & 0.45 \\
$_{68}$Er$_{99}^{167}$ & 6.44 & 97 & 0.46 \\
$_{68}$Er$_{100}^{168}$ & 7.77 & 96 & 0.46 \\
$_{69}$Tm$_{101}^{170}$ & 6.38 & 86 & 0.44 \\
$_{70}$Yb$_{102}^{172}$ & 8.14 & 74 & 0.43 \\
$_{70}$Yb$_{104}^{174}$ & 7.44 & 79 & 0.43 \\
$_{72}$Hf$_{106}^{178}$ & 7.62 & 64 & 0.41 \\
$_{73}$Ta$_{109}^{182}$ & 6.06 & 54 & 0.39 \\
$_{74}$W$_{109}^{183}$ & 6.19 & 58 & 0.40 \\
$_{74}$W$_{110}^{184}$ & 7.42 & 74 & 0.42 \\
$_{74}$W$_{111}^{185}$ & 5.75 & 64 & 0.41 \\
$_{74}$W$_{113}^{187}$ & 5.46 & 62 & 0.40 \\
$_{75}$Re$_{1131}^{188}$ & 6.24 & 55 & 0.39 \\
$_{75}$Re$_{113}^{188}$ & 5.73 & 55 & 0.39 \\
\hline
\end{tabular}
\end{table}

\begin{table}
\begin{tabular}{|c|c|c|c|}
\hline
Compound nucleus & $E_{max}$, MeV & $\bar{\Gamma}_\ga$, MeV &$T$, MeV \\
\hline
\multicolumn{4}{|c|}{Nonspherical actinides} \\
$_{90}$Th$_{143}^{233}$ & 4.96 & 21 & 0.30  \\
$_{91}$Pa$_{141}^{232}$ & 5.52 & 44 & 0.33  \\
$_{91}$Pa$_{143}^{234}$ & 5.12 & 48 & 0.35  \\
$_{92}$U$_{142}^{234}$ & 6.78 & 40 & 0.34  \\
$_{92}$U$_{143}^{235}$ & 5.27 & 25 & 0.31  \\
$_{92}$U$_{144}^{236}$ & 6.47 & 40 & 0.33  \\
$_{92}$U$_{145}^{237}$ & 5.30 & 29 & 0.31  \\
$_{92}$U$_{147}^{239}$ & 4.78 & 23 & 0.30  \\
$_{93}$Np$_{145}^{238}$ & 5.43 & 34 & 0.32  \\
$_{94}$Pu$_{146}^{240}$ & 6.46 & 40 & 0.33  \\
$_{94}$Pu$_{147}^{241}$ & 5.41 & 31 & 0.32 \\
$_{94}$Pu$_{148}^{242}$ & 6.22 & 37 & 0.33 \\
$_{95}$Am$_{147}^{242}$ & 5.48 & 42 & 0.33 \\
$_{95}$Am$_{149}^{244}$ & 5.29 & 50 & 0.35 \\
$_{96}$Cm$_{148}^{244}$ & 6.72 & 37 & 0.33 \\
$_{96}$Cm$_{149}^{245}$ & 5.70 & 39 & 0.33 \\
$_{96}$Cm$_{151}^{247}$ & 5.21 & 35 & 0.32 \\
\hline
\end{tabular}
\end{table}

Spherical nuclei possess a number of unique features that must be taken into
consideration (see the Introduction). However, the question of the applicability
to them of the formula (15) is at present difficult to answer categorically.
Indeed, nuclei of this sort apparently undergo a phase transition to the ``normal",
nonspherical state at temperatures
$$
T\sim \Delta\eps',
$$
where $\Delta\eps'$ is some characteristic width of the diffuse zone of the Fermi
distribution, a zone which owes its existence to the residual interaction. Meanwhile,
the spectrum of the emitted quanta (it is given by the integrand on the left-hand side
of (12)) is such that the energy averaged over it is equal to
$$
\bar{\eps}\cong 5T
$$
(see also [1], formulae (5) and (6)). Thus, many of the radiative transitions can,
roughly speaking, elude that region of the statistical distribution of the quasiparticles
where the distribution differs significantly from (9). Therefore, the attempts to apply
the formula (15) also to spherical nuclei, though not rigorous, is nevertheless of some
interest. It is natural to suppose that spherical nuclei have higher temperatures (and,
consequently, relatively low entropies; see also [1]). Comparison with the data on the
radiative widths apparently corroborates this trend. For example, for the compound
nucleus $_{79}$Au$_{119}^{198}$ we obtain $T = 0.45$ MeV, in the case of
$_{80}$Hg$_{122}^{202}$ we have $T = 0.56$ MeV, and, finally, $T = 0.62$ MeV for
$_{81}$Tl$_{123}^{204}$.

The experimental study of radiative-width fluctuations became possible only recently
as a result of an increase in the accuracy of their measurement, and comparison of the
theoretical formulas with experiment meets for the present with certain practical
difficulties. Let us discuss three specific nuclei, for which a selection of resonance
levels with accurately measured radiative widths nevertheless allowed the estimation
of the relaxation time $\tau$ from the formula (29) (see also (22) and (23)). Data on
two gadolinium isotopes are given in [11]; we took into consideration only the levels
for which the error in the radiative width is less than $10^{-2}$ eV. In the case of
$_{64}$Gd$_{92}^{156}$ (10 levels) $\overline{\Gamma}_\ga = 0.11$ eV,
$[\overline{(\Delta\Gamma_\ga)^2}]^{1/2} = 0.016$ eV, and $\hbar/\tau = 26$ eV. For
$_{64}$Gd$_{94}^{158}$ (11 levels) we have $\overline{\Gamma}_\ga = 0.089$ eV,
$[\overline{(\Delta\Gamma_\ga)^2}]^{1/2}  = 0.0087$ eV, and $\hbar/\tau  = 52$
eV. Let us also give the results of a similar analysis of the data on holmium [12]:
$_{87}$Ho$_{99}^{166}$ (21 levels) $\overline{\Gamma}_\ga = 0.091$ eV,
$[\overline{(\Delta\Gamma_\ga)^2}]^{1/2} = 0.0099$ eV, and  $\hbar/\tau = 39$ eV. Thus,
as far as we can judge, $\hbar/\tau \sim 50$ eV and $\tau\sim 10^{-17}$ sec, which
is a remarkably long time on the nuclear scale. We must, however, not forget that the
longest of the relaxation times $\tau$ of the system enters into the thermodynamic
theory (see formula (26)). The ``partial equilibrium" at each moment of time was
understood to have been established over the significantly shorter relaxation times
\footnote{From the data of the recent paper [13] we find that $\hbar/\tau\sim 20$ eV for the
compound nuclei Th$^{233}$ and U$^{239}$.}.

\section{Discussion. Is there enough time for the establishment of thermal equilibrium
in a nucleus?}

The question of relaxation in nuclear matter is of considerable interest. Thus far,
as far as we know, it has not been possible to estimate the characteristic time of
this process on the basis of any direct analysis
of the experimental data. Therefore, the observed fluctuations in the radiative
widths can be a valuable source of such information, and even the preliminary,
tentative figures ($\tau\sim 10^{-17}$ sec; see the preceding section) need to be
discussed. For the above-mentioned particular cases of resonance excitation by neutrons
of energy $< 1$ keV, the condition (24) was satisfied with three orders of magnitude
to spare---in other words, total thermal equilibrium was attained in the nucleus.
The situation can, however, change when we go over to higher kinetic energies of the
bombarding particles (see below).

It would, apparently, be somewhat naive to regard the time $t_1 = \hbar/\eps_0\sim 10^{-22}$
sec of transit of a quasiparticle through the nucleus as an estimate for the relaxation time.
Indeed, for example, the system is not in the least drawn nearer to the state of thermal
equilibrium by a coherent, reversible, purely elastic act of collision between a
quasiparticle and the surface of the nucleus
\footnote{Besides, a characteristic time $\sim t_1$ is quite capable of playing an important
role at the earliest stage of the development of the nuclear reaction. Here, however,
we are discussing only the late, final phase of the thermal-equilibrium establishment
process---see the end of the preceding section. In particular, it is extremely doubtful
that there will, at such times, remain reasonable physical criteria for distinguishing
the initial bombarding particle (or the corresponding quasiparticle).}. On the other hand,
inter-quasiparticle collisions appear to be quite an effective relaxation mechanism; it may
well turn out to be the dominant mechanism.

For obvious reasons, the thermal-radiation data used in the present paper actually pertained
to an extremely narrow part of the energy spectrum of the system. Let us now qualitatively
consider how the real conditions under which the relaxation process proceeds in the nucleus
should change upon further increase in the excitation energy of the nucleus. The neutron
width $\Gamma_n$ increases first in proportion to the square root of the distance from the
neutron-detachment threshold; then there we come into the stage of much more rapid exponential
growth. It is well known that owing to this phenomenon the resonance levels merge, forming
a continuous spectrum. But then whether the lifetime $\hbar/\Gamma$ of the compound nucleus
will be long enough for the establishment of total thermal equilibrium in the nucleus may
become doubtful, beginning from \footnote{It is possible that the so-called Erickson
fluctuations in the nucleon-nucleus interaction cross sections [14] are also due
to this circumstance.}
$$
\Gamma\cong\Gamma_n\geq 100\,\, \mathrm{eV}.
$$

In this connection, it is desirable to try critically reinterpret the method, based
on the neutron ``evaporation" process, for determining nuclear temperatures. It is
difficult for the present to judge how the fact that the state of the neutron-emitting
nucleus is not a totally equilibrium state will influence such an analysis. Not much
doubt has thus far been expressed about the evaporation temperatures probably because
their order of magnitude is quite plausible (and, in so far as we can judge, indeed
correct). However, as the experimental investigation of the reactions $(n, n')$ goes on,
attention will have to be paid not only to the absolute figures, but also to the behavior
of the relevant quantities. Of special interest, in particular, is the case when the
temperature of the compound nucleus as a function of its excitation energy is an almost
horizontal, non-monotonic in detail, and often simply a decreasing function. Unfortunately,
the authors of the corresponding publications give this remarkable circumstance
comparatively little consideration (see, for example [15]). We could have attempted to
interpret the decrease of the temperature with increasing excitation energy as some
giant random fluctuation, but it would have been difficult to conceive it as a phenomenon
that would occur with any degree of consistency. On the average, however, negative
specific heat is impossible for the nucleus. A state with negative specific heat is
totally unstable, and cannot be realized in nature [3].

We express our thanks to A.I. Baz', V.K. Voitovetskii, I.I. Gurevich, A.G. Zelenkov,
L.P. Kudrin, V.M. Kulakov, A.A. Ogloblin, I.M. Pavlichenkov, N.M. Polievktov-Nikoladze,
and K.A. Ter-Martirosyan for a discussion of the results of the paper.

\setcounter{equation}{0}

\renewcommand{\theequation}{A.\arabic{equation}}

\section*{Appendix}

Let us expound in some detail the integration over the boson energies in the formula (22).
The terms of the integral $J$ are identical in form, but individually each of them
contains a pole at $x_1 = x_2$. Therefore, it is sufficient to evaluate any of these
integrals in the principal value sense:
\begin{equation}\label{a-1}
    J=2\int_0^\infty dx_1\cdot x_1^4 e^{x_1}\coth\frac{x_1}2\,\mathrm{P}\!\int_0^\infty
    \frac{x_2^3dx_2}{(e^{x_1}-e^{x_2})(e^{x_1}-e^{-x_2})}.
\end{equation}
Let us transform the inner integral with the aid of the substitution $y = e^{-x_2}$,
representing it as a derivative with respect to some parameter:
\begin{equation}\label{a-2}
\begin{split}
    \mathrm{P}\!\!\int_0^\infty\frac{x_2^3dx_2}{(e^{x_1}-e^{x_2})(e^{x_1}-e^{-x_2})}&=
    e^{-x}\,\mathrm{P}\!\!\int_0^1\frac{(\ln y)^3dy}{(y-e^x)(y-e^{-x})}\\
    &=e^{-x}\frac{\partial^3}{\partial\nu^3}\,\mathrm{P}\!\!\int_0^1\frac{y^\nu dy}
    {(y-e^x)(y-e^{-x})},\quad \nu\to 0
    \end{split}
\end{equation}
($x = x_1$). In the decomposition
\begin{equation}\label{a-3}
    \mathrm{P}\!\int_0^1\frac{y^\nu dy}
    {(y-e^x)(y-e^{-x})}=\frac1{2\sinh x}\left\{J_\nu^+(x)-J_\nu^-(x)\right\},
\end{equation}
in which the integrand is expressed in partial fractions, the integrals
\begin{equation}\label{a-4}
    J_\nu^+(x)=\mathrm{P}\!\!\int_0^1\frac{y^\nu dy}{y-e^x},\quad
    J_\nu^-(x)=\mathrm{P}\!\!\int_0^1\frac{y^\nu dy}{y-e^{-x}}
\end{equation}
can conveniently be expressed as series. For this purpose, let us represent the
fractions by the corresponding geometric progressions, and let us also take into
consideration the formula
$$
\sum_{n=-\infty}^\infty\frac1{n+\nu}=\pi\cot\pi\nu.
$$
We then have
\begin{equation}\label{a-5}
    J_\nu^+(x)=\sum_{n=1}^\infty\frac{e^{-nx}}{n+\nu},\quad
    J_\nu^-(x)=-e^{-\nu x}\pi\cot\pi\nu+\frac1\nu-\sum_{n=1}^\infty
    \frac{e^{-nx}}{n-\nu}.
\end{equation}
Taking the limit in accordance with (A.2), we obtain
\begin{equation}\label{a-6}
    J=192\int_0^\infty\left\{\frac{\xi^8}3-\frac{\pi^2}3\xi^6-\frac{\pi^4}{90}
    \xi^4\right\}\frac{d\xi}{\sinh^2\xi}+6\sum_{n=1}^\infty\frac1{n^4}
    \int_0^\infty\frac{\xi^4e^{-n\xi}}{\sinh^2(\xi/2)}d\xi.
\end{equation}
It is easy to see that the integral standing under the summation sign
\begin{equation}\label{a-7}
    \int_0^\infty\frac{\xi^4e^{-n\xi}}{\sinh^2(\xi/2)}d\xi=4\int_0^1
    \frac{(\ln y)^4y^n}{(y-1)^2}dy=4\frac{\partial^4}{\partial\nu^4}\,
    \mathrm{P}\!\!\int_0^1\frac{y^{n+\nu}dy}{(y-e^x)(y-e^{-x})},\quad \nu\to 0
\end{equation}
is essentially of the same type as (A.2), except that it is differentiated once
more with respect to $\nu$ and that the additional passage to the limit $x\to 0$
will also be necessary. The subsequent simple computations, besides the passages
to the limit, also include convenient re-designations of the indices of the double
summation. As a result, we obtain the formula (27).


\begin{thebibliography}{99}

\bibitem{1} V.G. Nosov and A. M. Kamchatnov, Zh. Eksp. Teor.
Fiz. {\bf 65,} 12 (1973) [Sov. Phys.-JETP {\bf 38,} 6 (1974)]; arXiv nucl-th/0311045.

\bibitem{2} L.D. Landau, Zh. Eksp. Teor. Fiz. {\bf 30,} 1058 (1956)
[Sov. Phys.-JETP {\bf 3,} 920 (1957)].

\bibitem{3} L.D. Landau and E.M. Lifshitz, Statisticheskaya
fizika (Statistical Physics), Fizmatgiz, 1964 (Eng.
Transl., Addison-Wesley Publ. Co., Reading, Mass.,
1969).

\bibitem{4} V.G. Nosov, Zh. Eksp. Teor. Fiz. {\bf 57,} 1765 (1969)
[Sov. Phys.-JETP {\bf 30,} 954 (1970)].

\bibitem{5} A.M. Kamchatnov and V.G. Nosov, Zh. Eksp. Teor.
Fiz. {\bf 63,} 1961 (1972) [Sov. Phys.-JETP 36, 1036 (1973)]; arXiv nucl-th/0311011.

\bibitem{6} V.B. Berestetskii, E.M. Lifshitz, and L.P. Pitaevskii, Relyativistskaya
kvantovaya teoriya (Relativistic Quantum Theory), Part. l. Nauka, 1968 (Engl. Transl.,
Pergamon, New York, 1971).

\bibitem{7} L.D. Landau and E.M. Lifshitz, Kvantovaya mekhanika (Quantum Mechanics),
Fizmatgiz, 1963 (Eng. Transl., Addison-Wesley, Reading, Mass., 1965).

\bibitem{8} L.D. Landau and E.M. Lifshitz, Teoriya polya (The Classical Theory of
Fields), 2nd ed., Goztekhizdat, 1948, (549, pp. 134-135 (Eng. Transl., Addison-Wesley,
Reading, Mass., 1962).

\bibitem{9} M.D. Goldberg, S.F. Mughabghab, S.N. Purohit, B.A. Magurno, and V.M. May,
Neutron Cross Sections, Vol. IIC, Brookhaven National Laboratory, 1966.

\bibitem{10} J.R. Stehn, M.D. Goldberg, R. Wiener-Chasman, S.F. Mughabghab,
B.A. Magurno, and V.M. May, Neutron Cross Sections, Vol. Ill, Brookhaven National
Laboratory, 1965.

\bibitem{11} S.J. Friesenhahn, M.P. Fricke, D.G. Costello, W.M. Lopez, and A.D. Carlson,
Nucl. Phys. {\bf A146,} 337 (1970).

\bibitem{12} M. Asghar, C.M. Chaffey, and M.C. Moxon, Nucl. Phys. {\bf A108,} 535 (1968).

\bibitem{13} F. Rahn, H.S. Camarda, G. Hacken, W.W. Havens, Jr.,
H. I. Liou, J. Rainwater, M. Slagowitz, and S. Wynchank,
Phys. Rev. {\bf C6,} 1854 (1972).

\bibitem{14} T. Erickson and T. Mayer-Kuckuk, Ann. Rev. Nuc. Sci.
Vol. {\bf 16,} 1966 [Russ. transl. Usp. Fiz. Nauk {\bf 92,} 271 (1967)].

\bibitem{15} R.O. Owens and J.H. Towle, Nucl. Phys. {\bf A112,} 337
(1968).

\end{thebibliography}
\end{document}